\documentstyle[aps,pre,preprint,psfig,rotate]{revtex}
\begin{document}
\draft
\tighten

\newcommand{\eql}[1]{\label{eq:#1}}
\newcommand{\eq}[1]{Eq.~(\ref{eq:#1})}
\newcommand{\dd}[2]{\frac{\partial #1}{\partial #2}}
\newcommand{\average}[1]{\left\langle #1 \right\rangle}
\newcommand{\saverage}[1]{\langle #1 \rangle}
\newcommand{\order}{{\cal O}}
\newcommand{\vect}[2]{\left(\begin{array}{c} #1 \\#2 \end{array}\right)}
\newcommand{\mb}[1]{\mbox{\boldmath $#1$}}
\newcommand{\tvector}{\mb{\theta}}
\newcommand{\fake}[2]{\left#2\begin{picture}(0,#1)\end{picture}\right.}
\newcommand{\xvector}{\mb{\xi}}
\newcommand{\split}[1]{}
\renewcommand{\split}[1]{\nonumber\\ && #1}

\title{
\vskip -100pt
{
\begin{normalsize}
\begin{flushright}
cond-mat/9905027\\
THU-99/10\\
July 1999\\
\end{flushright}
\vskip  70pt
\end{normalsize}
}
  Kinetic approach to the Gaussian thermostat in a
  dilute sheared gas in the thermodynamic limit
}

\author{R. van Zon\thanks{email: R.vanZon@phys.uu.nl}}
\address{Institute for Theoretical Physics,
University of Utrecht\\ Princetonplein 5, 3584 CC
Utrecht, The Netherlands}
\date{July 8, 1999}

\maketitle

\begin{abstract}

A dilute gas of particles with short range interactions is considered in a
shearing stationary state.  A Gaussian thermostat keeps the total kinetic
energy constant.  For infinitely many particles it is shown that the thermostat
becomes a friction force with  constant friction coefficient. For finite number
of  particles $N$, the fluctuations around this constant are of order $1/\sqrt
N$, and distributed approximately  Gaussian with deviations for large
fluctuations. These deviations prohibit a derivation of
fluctuation-dissipation relations far from equilibrium, based on the
Fluctuation Theorem.

\end{abstract}


The interest in the relation between non-equilibrium statistical mechanics and
microscopic equations of motion, which already occupied Boltzmann, has revived
in recent years, on the one hand due to the development of chaos theory, but
even more due to results from non-equilibrium molecular
dynamics\cite{Mareschal,Evans}.  The main focus in the field is on stationary
states.  A stationary state, if it is not the equilibrium state, is the result
of an external driving force. But this force performs work on the system, so it
heats up (viscous heating, Ohmic heating).  In simulations this is often
remedied by the introduction of a mechanical thermostat: one adds a friction
force, $-\alpha\vec v_i$, in the equation of motion for the velocity $\vec v_i$
of each particle $i$, to keep the energy constant.  For the thermostat variable
$\alpha$ there are several choices.  One could take it constant, but then one
only gets a constant energy on average. It is also possible to have $\alpha$
time dependent, such that the total kinetic energy is constant (iso-kinetic
Gaussian thermostat) or the total energy is constant (iso-energetic Gaussian
thermostat)\cite{Evans}.  Neither of these thermostats are very realistic, as
the dissipation of the heating would more likely occur at the boundary, where
the system is in contact with a heat bath, say.  Other boundary formulations
where the driving force and the thermostat are combined have also been
studied\cite{Chernov,Dellago4}. One hopes the choice of the thermostat doesn't
matter in the thermodynamic limit.  The equivalence of a constant $\alpha$
thermostat, the iso-kinetic thermostat and iso-energetic thermostat was
proposed by Gallavotti\cite{Gallavotti2}.

The extra term in the equations of motion destroys the Liouvillian character of
the flow:  a given volume in phase space will not retain that volume. As the
available phase space is usually finite, this means that on average over time
the volume either stays constant (conservative case) or contracts (dissipative
case).  In a dissipative system a stationary state can exist only on a course
grained scale, the dissipation continues forever but on ever finer scales.
This dissipation happens at a rate called the phase space contraction rate
which is proportional to the average of the thermostat variable $\alpha$.  This
rate can be identified\cite{Ruelle,Cohen2} with the irreversible entropy
production\cite{DeGrootMazur}.  If we make this identification with a physical
quantity, the precise implementation (iso-kinetic, iso-energetic, constant
$\alpha$, $\ldots$) of the thermostat should not influence the average of
$\alpha$ in the thermodynamic limit. Cohen\cite{CohenTransport} suggested that
a mechanical and a physical thermostat may give the same results as long as the
rate of heating is much less than the rate at which heat can be transported to
the wall and absorbed there.  This suggests that when the rate of heating
becomes to large, the thermostat does make a difference.  At that point one
also expects the assumption of local equilibrium underlying non-equilibrium
thermodynamics to break down, and the entropy production may no longer have the
form that was used to identify it with $\alpha$.

We want to know which thermostat to use for analytic treatment of dilute gases
in non-equilibrium stationary states.  As we are interested in the limit of
many particles, having to use an $\alpha$ dependent on all these particles
would certainly make work more difficult.  In other analytic work on
non-equilibrium stationary states, one simply takes a constant
$\alpha$\cite{LeeDufty,Lutsko}.   A sketch of a proof of the equivalence  of a
Gaussian iso-kinetic thermostat, a Gaussian iso-energetic thermostat and a
Nos\'e-Hoover thermostat, was already given  by Evans and
Sarman\cite{EvansSarman}. In this paper, we will demonstrate the equivalence of
an iso-kinetic Gaussian thermostat and a constant thermostat in the
thermodynamic limit  using kinetic theory on the Boltzmann level (i.e.\ at low
densities) for a sheared system of short range interacting particles.

\section{Sheared gas with SLLOD}

\begin{figure}[th]
 \centerline{\psfig{file=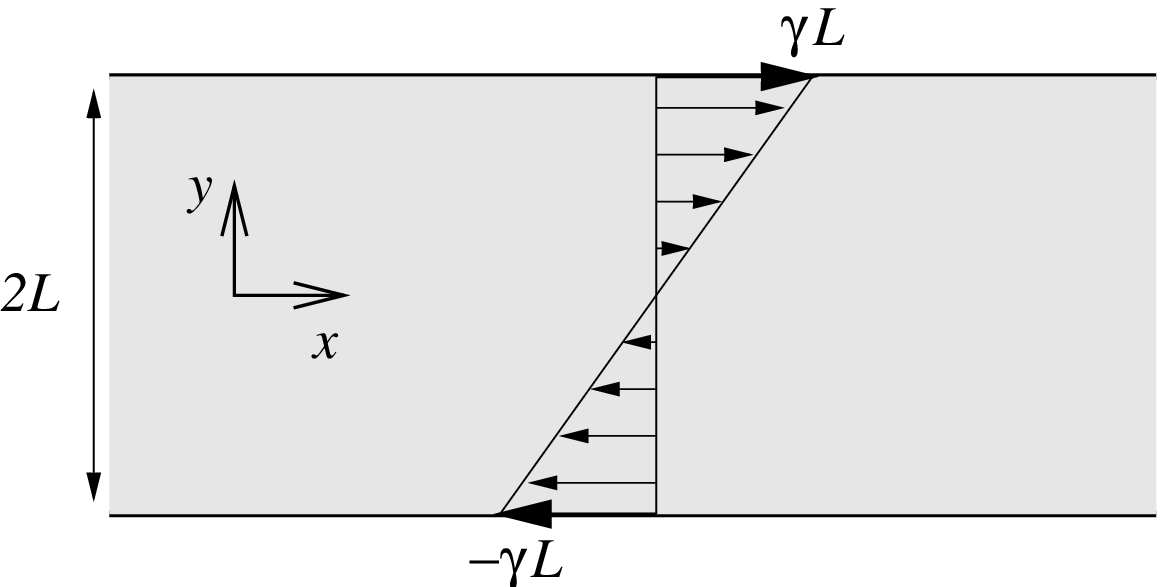,height=3.5cm}}
\caption{Velocity profile in a gas under shear}
\label{fig:shear}
\end{figure}

We consider a dilute gas of $N$ particles in $d$ dimensions, under shear: the
gas is contained between two plates a distance $2L$ apart
(Fig.~\ref{fig:shear}).  The two plates are moving in opposite directions with
a velocity $\gamma L$.   For not too large $\gamma$ one expects that a linear
velocity profile will develop, so that the fluid velocity at $y$ is $\gamma
y\hat{x}$.

We are interested in bulk properties, so we let $L$ go to infinity, while the
shear rate $\gamma$ and the density $\rho$ are fixed.  To show the equivalence
of the constant  $\alpha$ thermostat and the Gaussian thermostat, it would in
principle  suffice just to take  the horizontal dimensions infinite, but this
way we can also move the boundary conditions to infinity.  In the real physical
system, as $L$ get larger, the laminar flow becomes  unstable and the system
eventually develops turbulence. However, the  thermostats we will consider
assume the stability of the laminar  flow\cite{Evans}, and suppress turbulence.

There is a well known and often used set of equations for molecular simulations
that describe shear, the SLLOD equations\cite{Evans}:
\begin{eqnarray}
 \dot{\vec q}_i &=& \vec p_i + \gamma y_{i} \hat{x}\eql{S1}
\\
 \dot{\vec p}_i &=& \vec F_i - \gamma p_{iy} \hat{x}
		- \alpha \vec p_i,\eql{S2}
\end{eqnarray}
in which $\vec q_i,\vec p_i$ are the phase space coordinates of particle $i$,
$\alpha$ is the thermostat variable and $\vec F_i$ represents the forces
between the particles.  The mass of the particles is taken to be one. The shear
rate is constant, but $\alpha$ is not.  It is constructed such that the kinetic
energy $K=\sum_i\|\vec p_i\|^2/2$ in the system is exactly constant, which
gives
\begin{eqnarray*}
	\alpha = \frac{1}{2K}
\sum_{i=1}^{N}\left(\vec F_i\cdot\vec p_i - \gamma
p_{ix}p_{iy}\right).
\end{eqnarray*}
This is the iso-kinetic Gaussian thermostat. Note that $\alpha$ depends on the
positions and momenta of all the particles.

The interpretation of \eq{S1} is that $\vec p_i$ is the peculiar velocity of
particle $i$ with respect to the local fluid element that has velocity $\gamma
y\hat{x}$. In the laboratory frame, a reasonable set of equations to write down
is
\begin{eqnarray*}
	\dot{\vec q}_i = \vec v_i, &\;\; &
	\dot{\vec v}_i = \vec F_i
		- \alpha \left(\vec v_i-\gamma y_i\hat{x}\right).
\end{eqnarray*}
This particular form of the thermostat term is chosen because a linear velocity
profile is expected, and we want the temperature to be constant, so the kinetic
energy in the frame that moves with that local velocity is to be dissipated.  A
Boltzmann equation for the one-particle distribution function will give an
appropriate description at low densities. This equation has to be supplemented
by the boundary condition that the average velocities at the boundaries $y=\pm
L$ are $\pm\gamma L$.  To get rid of the $L$ dependence, one can transform the
velocities to peculiar velocities: $\vec p_i=\vec v_i-\gamma y_i \hat{x}$. The
average (peculiar) velocity now has to be zero at the boundaries, so when
$L\rightarrow\infty$ they have to be zero at infinity: this is the same
boundary condition as for the standard Boltzmann equation.  The transformation
to peculiar velocities yields the SLLOD equations:
\[
 \dot{\vec{q}_i} = \vec{v}_i = \vec{p}_i
  + \gamma y_i \hat{x},
\]
\[
 \dot{\vec{p}_i} = \dot{\vec{v}}_i
 -\gamma \dot{y}_i \hat{x}
 = \vec F_i - \alpha \vec p_i - \gamma p_{iy} \hat{x}.
\]

\section{Kinetic Approach}
\subsection{Effective motion of the thermostat}

The Gaussian thermostat involves an $\alpha$ which depends on the position and
velocity of every particle, and so it varies in time. We'll now derive
equations of motion for the thermostat for which we do not need to know all
these positions and velocities, by introducing one extra thermostat variable
$\beta$. The derivation is in two steps: first we consider free flight, and
then we take into account the effect of collisions.

During free flight $\vec F$ is zero so the thermostat is given by
\begin{equation}
  \alpha = \frac{1}{N} \sum_{i=1}^{N}
  \frac{-\gamma p_{iy}p_{ix}}{2K/N}
\eql{alpha}
\end{equation}
Using the equations of motions, one finds that the time derivative is
\begin{equation}
  \dot{\alpha} = -2\alpha^2 + \frac{\gamma^2}{2K}
                \sum_{i=1}^N p_{iy}^2 \equiv
                -2\alpha^2 + \beta
\eql{alphadot}
\end{equation}
where we've defined the last part as the second  thermostat variable $\beta$:
\begin{equation}
  \beta = \frac{1}{N} \sum_{i=1}^{N}
  \frac{\gamma^2p_{iy}^2}{2K/N}
\eql{beta}
\end{equation}
We combine these to the thermostat vector  $\tvector=(\alpha,\beta)$. The time
derivative of  $\beta$ is expressable again in terms of $\alpha$ and  $\beta$:
\begin{eqnarray}
  \dot{\beta} = -2\alpha\beta.
\eql{betadot}
\end{eqnarray}
There is a conserved quantity
\[
  H = \frac{\alpha^2-\beta}{2\beta^2},
\]
After change of variables to $X = 1/(2\beta)$ and $P =\alpha/\beta$, this
conserved quantity takes on a Hamiltonian form $H(X,P) = \frac{1}{2} P^2 - X$.
The equations of motion are $\dot{X} = P$ and $\dot{P} = 1$. The general
solution, transformed back to $\tvector$, has the form
\[
  \tvector(t) = \frac{1}{-2H+(t-t_0)^2} \vect{t-t_0}{1},
\]
with $t_0$ a constant.

So far, we only considered free flight. The duration of a free flight is very
small in a system of $N$ particles: it is of the order of $(N\nu)^{-1}$, where
$\nu$ is the collision frequency of a single particle.  In a collision,
$\tvector$ changes by an amount:
\begin{eqnarray*}
	\Delta \tvector &=&
        \frac{-\gamma}{2K}
	\vect{
	p'_{1x}p'_{1y}
	+p'_{2x}p'_{2y}
	-p_{1x}p_{1y}
	-p_{2x}p_{2y}
	}
	{-\gamma\left[
	p^{\prime 2}_{1y}
	+p^{\prime 2}_{2y}
	-p^2_{1y}
	-p^2_{2y}
	\right] }\nonumber\\
 &\equiv& \frac{\xvector(\vec{p}_1,\vec{p}_2,\hat{n})}{N},
\end{eqnarray*}
where primes denote the value of the variables after collisions and $\hat{n}$
is the collision parameter, i.e.\ $\hat{n}=(\vec{p}'_{21}-\vec{p}_{21})/
\|\vec{p}'_{21}-\vec{p}_{21}\|$. As $K$ is of order $N$, $\Delta\tvector$ is of
order $N^{-1}$. $\tvector$ itself is of order one. The definition makes
$\xvector$ of order one.  During a typical free flight time of one particle,
every particle in the system has collided once on average. In each collision
two particles are involved, so $N/2$ collisions have taken place in the system
during this time. Thus, during one free flight time, the thermostat got $N/2$
changes of order $N^{-1}$ and this adds up to an effect of order one, because,
as we will see, the average of $\Delta\tvector$ is non-zero.

We will see later that the effect of the thermostat depending on all the
particles -- an unphysical idea in some sense -- results just in a fluctuating
thermostat with well defined mean and distribution. We are interested in this
distribution. It will depend on the distribution of $\Delta \tvector$, which
depends on the velocity distribution in the system, which in turn is affected
by the thermostat distribution. But some general properties can already be
derived without this subtle interplay.

We want to write down a Boltzmann equation for the  probability distribution
function $F(\tvector;t)$ of the  thermostat:
\begin{equation}
  \dd{F}{t} + \dd{}{\tvector}\cdot(\dot{\tvector} F)
  = \left. \dd{F}{t}\right|_{c}.
\eql{generic}
\end{equation}
Here $\dot{\tvector}$ is given by \eq{alphadot} and \eq{betadot}. The collision
integral counts the number of states that are lost and gained in collisions:
\begin{eqnarray*}
  &&\left. \dd{F}{t}\right|_{c}
  = \int d\tvector^* d\vec p_1 d\vec p_2
	P(\vec{p}_1,\vec{p}_2,\tvector^*;t)
	\int d\hat{n}\;\split{\quad}
  \mbox{rate of $(\vec{p}_1,\vec{p}_2,\hat{n})$}
  \left\{ \delta\left(\tvector-\tvector^*-\frac{\vec{
   \xi}}{N}\right)
   -\delta\left(\tvector-\tvector^*\right) \right\},
\end{eqnarray*}
in which $P$ is the joint distribution of $\vec  p_1$, $\vec p_2$ and
$\tvector$, and
\begin{eqnarray*}
  \mbox{rate of $(\vec{p}_1,\vec{p}_2,\hat{n})$} &=&
  \frac{N}{2} \rho B(\hat{n},\vec{p}_{21})
\end{eqnarray*}
where $\vec{p}_{21}=\vec{p}_2-\vec{p}_1$. $B$ is the rate of collisions with
$\hat{n}$ given that the colliding particles have relative velocity
$\vec{p}_{21}$. It can be expressed in the differential cross section
$s(\hat{n},\vec{p}_{21})$, which measures the number of deflected particles per
unit solid angle around $\vec{p}'_{21}$ when a beam of particles of unit
density is incident on one other particle. In $d$ dimensions:
\[
 B(\hat{n},\vec{p}_{21})= \|\vec{p}_{21}\|
s(\hat{n},\vec{p}_{21})
 2^{d-1} \left|\hat{n}\cdot\hat{p}_{21}\right|^{d-2}
\]
where $\hat{p}_{21}=\vec{p}_{21}/\|\vec{p}_{21}\|$. The last factor is the
Jacobian that arises because we integrate over $\hat{n}$ while $s$ is defined
per unit solid angle of $\vec{p}'_{21}$. Strictly, we ought to take
$\vec{p}_2-\vec{p}_1+\gamma\hat{x}(y_2-y_1)$ instead of $\vec{p}_{21}$, but in
the Boltzmann approach the two particles are taken at the same position when
they collide. In an Enskog-type approach this would matter.

To proceed, we need an expression for $P$ in terms of $F$ and the one particle
distribution function $f(\vec p)$, which we will take normalized to one. For
$f$, we can also write down a Boltzmann equation, so we will have a system of
two coupled Boltzmann equations for $f$ and $F$. To derive the standard
Boltzmann equation\cite{Cercignani} for $f$, one uses the Stosszahlansatz that
states that the two-particle distribution function $f_2(\vec{p}_1,\vec{p}_2)$
is proportional to the product of the one particle distribution functions
$f(\vec{p}_1)f(\vec{p}_2)$, i.e.\ that the two particles are uncorrelated when
they collide. We want a generalization of the Stosszahlansatz for $P$, but
setting $P(\vec p_1,\vec p_2,\tvector)=f(\vec p_1) f(\vec p_2) F(\tvector)$
can't quite be right for the following reason.

The Stosszahlansatz can be generated by taking the phase space density
$\rho(\{\vec p_i\})$ to be $\prod_{i=1}^{N} f(\vec p_i)$, i.e.\ all the
particles are uncorrelated (arguably, this is too strong a condition, but it
will serve to make our point).  Let
\[
 \tvector_s (\vec p)= \frac{N}{2K}\vect{-\gamma p_y
 p_x}{\gamma^2p_y^2},
\]
such that
\[
  \tvector = \frac{1}{N}\sum_{i=1}^{N}
  \tvector_s(\vec p_i)
\]
then the quantity
\begin{eqnarray*}
  \Pi (\vec p_1, \vec p_2, \tvector) &\equiv& \int
  d\vec p_3 \ldots d\vec p_N \rho(\{ \vec p_i\})
  \delta\left( \tvector - \frac{\sum_{i=3}^{N}
  \tvector_s(\vec p_i)}{N-2}\right)\\ &
  \equiv & f(\vec p_1) f(\vec p_2) \Phi (\tvector)
\end{eqnarray*}
factorizes, because the delta function doesn't involve  $\vec p_1$ and $\vec
p_2$. One easily derives that
\[
  P(\vec p_1,\vec p_2,\tvector) =
 \left\{\frac{N}{N-2}\right\}^2\Pi\left(\vec p_1, \vec
 p_2, \frac{N\tvector-{\tvector}_s(\vec
 p_1)-\tvector_s(\vec p_2)}{N-2}\right)
\]
and this doesn't factorize.

$P$ does however factorize to zeroth order when we expand in $N^{-1}$. To see
this, we first expand the expression for $P$ in terms of the factorized $\Pi$:
\begin{eqnarray*}
 && P(\vec p_1,\vec p_2,\tvector) =
 f(\vec p_1)f(\vec p_2)\fake{16}{\{}
  \Phi(\tvector)\left[1+\frac{4}{N}\right] +
  \split{\quad +}
 \frac{1}{N}
 \left[2\tvector-\tvector_s(\vec p_1)-\tvector_s(\vec
 p_2)\right]\cdot\dd{\Phi}{\tvector}  \fake{16}{\}}
 +\order(N^{-2}).
\end{eqnarray*}
$F(\tvector)$ is given by $\int d\vec p_1 d\vec p_2 P(\vec p_1, \vec p_2,
\tvector)$ so integrating the above formula gives a relation between $F$ and
$\Phi$:
\[
 F(\tvector) = \Phi(\tvector) +\frac{1}{N} \left\{
 4\Phi(\tvector)+2\left[\tvector
 -\langle\tvector_s\rangle\right]\cdot\dd{\Phi}{\tvector}
 \right\} +\order(N^{-2}),
\]
where $\saverage{\tvector_s}=\int d\vec p f(\vec p)   \tvector_s(\vec p)$. This
relation can be inverted  up to order $N^{-1}$:
\[
 \Phi(\tvector) = F(\tvector) -\frac{1}{N} \left\{
 4F(\tvector)+2\left[\tvector-
 \langle\tvector_s\rangle\right]\cdot\dd{F}{\tvector}
 \right\}
 +\order(N^{-2}).
\]
When we put this back into the the formula for $P$,  we get $P$ expressed in
terms of $f$ and $F$:
\begin{eqnarray*}
 &&  P(\vec p_1, \vec p_2,\tvector)  =
 f(\vec p_1)f(\vec p_2)\fake{16}{\{}
  F(\tvector) + \split{\quad +}
 \frac{1}{N} \left[2\saverage{\tvector_s}-\tvector_s(\vec
 p_1)-\tvector_s(\vec p_2)\right]\cdot\dd{F}{\tvector}
 \fake{16}{\}} +\order(N^{-2}) .
\end{eqnarray*}
This will serve as our Stosszahlansatz. We insert it in the collision integral
and perform the  $\tvector^*$ integration:
\begin{eqnarray}
   \left.\dd{F}{t}\right|_{c} &=&
 -\left\{\int d\vec{p}_1 d\vec{p}_2 d\hat{n}
 f(\vec{p}_1) f(\vec{p}_2)
 \rho
 B(\hat{n},\vec{p}_{21})\frac{\xvector}{2}\right\} \cdot
 \dd{F}{\tvector}\nonumber\\
 &+&\frac{1}{N}\fake{16}{\{}
 \int d\vec{p}_1  d\vec{p}_2 d\hat{n}
 f(\vec{p}_1) f(\vec{p}_2)
 \rho B(\hat{n},\vec{p}_{21})
 \frac{\xvector}{2} \times
 \split{\times}
 \left[\tvector_s(\vec{p}_1)+\tvector_s(\vec{p}_2)-
 2\langle\tvector_s\rangle \right]
 \fake{16}{\}}:\dd{^2F}{\tvector\partial\tvector},
 \eql{star}
\end{eqnarray}
where we expanded in $N^{-1}$ once more. We define the  collisional averages:
\[
   \vect{a}{b} \equiv
   \frac{1}{2}\int d\vec p_1 d\vec p_2
   d\hat{n} f(\vec p_1)f(\vec p_2) \rho
   B(\hat{n},\vec{p}_{21})
   \xvector(\vec p_1,\vec p_2,\hat{n}).
\]
The Boltzmann equation for the thermostat can now be  written as
\begin{eqnarray*}
&&
   \dd{F}{t} + \dd{}{\alpha}\left\{
   \left[-2\alpha^2+\beta+a\right] F \right\}
   + \dd{}{\beta}\left\{\left[ -2\alpha\beta+b\right]
   F\right\}
   \split{\quad}\;=\;
   \order(N^{-1}).
\end{eqnarray*}

\noindent This is of the form of \eq{generic} in which points in the $\tvector$
phase space follow the effective free flight dynamics
\[
	\dot{\tvector} = \vect{-2\alpha^2 + \beta +
 a}{-2\alpha\beta+b} .
\]
This amounts to just adding the average effect of  collisions to the change of
$\tvector$.

\begin{figure}
 \centerline{\psfig{file=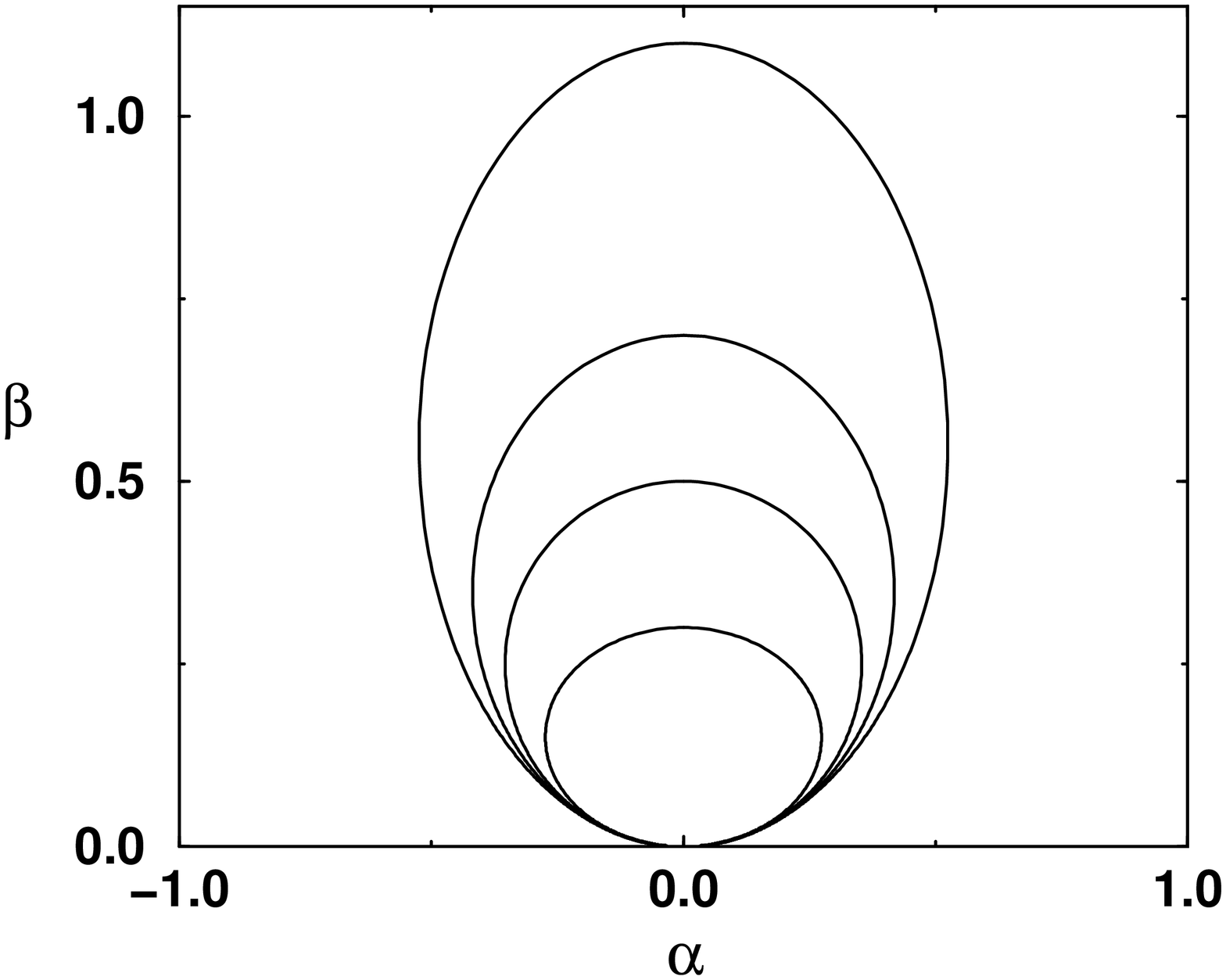,angle=-90,width=7cm}}
\centerline{\psfig{file=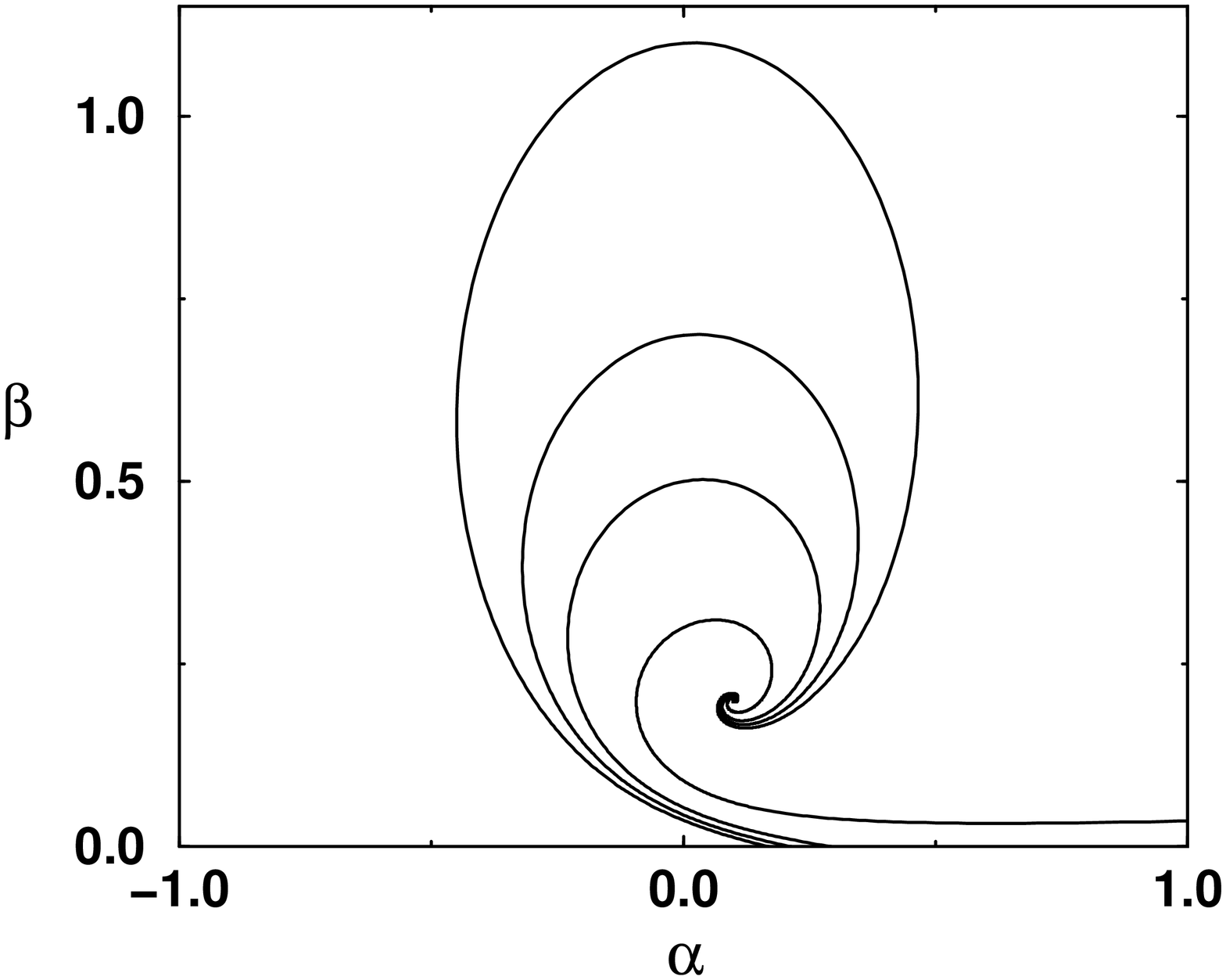,angle=-90,width=7cm}}
\caption{Some typical $\tvector$ trajectories of the
thermostat. The first figure is without collisions, the
second one is with collisional averages $a$ and $b$ set to
the arbitrary values of $-0.18$ and $0.04$,
respectively.}
\label{fig:phasespace}
\end{figure}

Some typical trajectories in $\tvector$ phase space are plotted in
Fig.~\ref{fig:phasespace}, both with and without the effect of collisions.
When $(a,b)\neq 0$, it is no longer possible to derive the equations of motion
from a Hamiltonian: the motion is dissipative, and there exists a fixed point
$\tvector_0=(\alpha_{0},\beta_{0})$, defined by
\begin{equation}
 -2\alpha_{0}^2+\beta_{0}+a = 0 \; ;\; 2
 \alpha_{0} \beta_{0} =b .
\eql{fixed}
\end{equation}
Physically, one expects the system to heat up without a thermostat, so the
thermostat should act as a friction: $\alpha_{0}$ is positive.  To consider the
stability of the fixed point, we linearize the equation of motion around it.
Writing $\tvector=\tvector_0+\delta\tvector$, we get
\begin{equation}
 \delta\dot{\tvector} =
  -\left(\begin{array}{cc}
    4\alpha_{0} & -1 \\
    2\beta_{0} & 2\alpha_{0}
  \end{array}\right)
 \delta\tvector .
\eql{matrixL}
\end{equation}
The eigenvalues of the matrix are $3\alpha_{0} \pm
\sqrt{\alpha_{0}^2-2\beta_{0}}$. $\beta_{0}$ is positive, so the fixed point is
stable only if $\alpha_{0}>0$, consistent with the physical expectation.
Furthermore, from the definitions in \eq{alpha} and \eq{beta}, one can see that
$\alpha_{0}^2 < \beta_{0}$, so the eigenvalues are complex and the fixed point
is a stable spiral. If we take an initial distribution within the domain of
attraction, all points will end up in this fixed point.  Thus, in the
stationary state, the distribution of the thermostat is a delta function at
$\tvector_0$:
\[
 F(\tvector)=\delta(\tvector-\tvector_0)
\]

\subsection{Boltzmann equation for the one particle  distribution function}

As we mentioned, the analysis is not complete without a second Boltzmann
equation, for the one particle distribution function $f(\vec p;t)$. The
difficult part a priori is what to do with the free flight term $-\alpha\vec
p_i$. The analysis of the thermostat variables however showed that the $\alpha$
is, in the stationary case, a constant, so we just replace $\alpha$ by
$\alpha_0$:
\[
 \dd{f}{t} + \dd{}{\vec p}\left\{[-\gamma
 p_y\hat{x}-\alpha_0\vec p] f\right\} = J(f,f),
\]
where we considered again only the uniform case. For the stationary case we are
interested in, also the time derivative disappears. The collision integral
is\cite{Cercignani}
\begin{eqnarray*}
 &&	\left. J(f,f)\right|_{\vec p=\vec p_1} = \int
 d\vec{p}_2 d\hat{n}
 \rho B(\hat{n},\vec{p}_{21}) \times
 \split{\quad\times}
 \left\{ f(\vec{p}_1')
 f(\vec{p}_2') - f (\vec{p}_1) f(\vec{p}_2) \right\} .
\end{eqnarray*}
In some analytic work\cite{LeeDufty,Lutsko}, a constant $\alpha$ is also used
and its value is determined by demanding that the second moment
$\saverage{\|\vec p\|^2}=\int \|\vec{p}\|^2 f(\vec p)\,d\vec p$ is a conserved
quantity. This yields
\begin{equation}
 \alpha = - \gamma \frac{ \average{p_x p_y} }{
 \average{\|\vec  p\|^2}} .
\eql{aav}
\end{equation}
Note that $\average{\|\vec p\|^2}$ is constant and equal to $2K/N$, so this
choice of $\alpha$ is very analogous to \eq{alpha}. In fact the solution of
\eq{fixed} determining the thermostat fixed point, is this $\alpha$, together
with
\begin{equation}
 \beta = \gamma^2 \frac{ \average{p_y^2} }{
 \average{\|\vec  p\|^2} }.
\eql{bav}
\end{equation}
To show this we consider $a$ and $b$:
\begin{eqnarray*}
 a &=& \frac{-\gamma}{2K/N} \int d\vec p_1 d\vec p_2
 d\hat n \,
 \rho B(\hat{n},\vec{p}_{21}) f(\vec p_1)f(\vec p_2)
 \times\split{\times}
  \frac{1}{2}\left\{p'_{1x}p'_{1y}+p'_{2x}p'_{2y}
 -p_{1x}p_{1y}-p_{2x}p_{2y}\right\} \\
 &=&\frac{-\gamma}{\average{\|\vec p\|^2}} \int d\vec p_1
 d\vec p_2 d\hat n\,
 \rho B(\hat{n},\vec{p}_{21})  p_{1x} p_{1y} \times
 \split{\times}
 \left\{ f (\vec{p}_1')
 f(\vec{p}_2) - f (\vec{p}_1) f(\vec{p}_2) \right\} \\
 &=& \frac{-\gamma}{\average{\|\vec p\|^2}}\int d\vec p
 \,J(f,f) p_x p_y ,\\
 b &=& \frac{\gamma^2}{\average{\|\vec p\|^2}}\int d\vec p
 \,J(f,f) p_y^2 .
\end{eqnarray*}
We insert for $J(f,f)$ the left hand side of the  Boltzmann equation and find
after partial integration
\[
 a = 2\alpha_0
 \frac{\average{p_xp_y}}{\average{\|\vec p\|^2}}
 - \gamma^2 \frac{\average{p_y^2}}{\average{\|\vec
 p\|^2}} = 2\alpha_0\alpha-\beta
\]
\[
 b= 2\alpha_0\gamma^2\frac{\average{p_y^2}}
 {\average{\|\vec p\|^2}}
 = 2\alpha_0\beta
\]
Combined with \eq{fixed}, we see that indeed   $(\alpha_0,\beta_0)$ $=$
$(\alpha,\beta)$ given by \eq{aav}  and \eq{bav}, is the self-consistent
solution.

\section{Thermostat fluctuations}

\label{sec:FD}

When $N$ is finite, there are fluctuations around the thermostat value
$\alpha_0$. One can see that the first correction, on the right hand side of
\eq{star} has a diffusive form, with a diffusion coefficient of order $N^{-1}$.
Combined with the drift towards a fixed point from the left hand side, this
will make the thermostat distribution $F$ peaked around this point, with a
width of order $N^{-1/2}$. For large $N$ the width is so small that the
linearized equation, \eq{matrixL} is a good approximation for most of the
distribution.  For a linear fixed point, the distribution would be Gaussian. So
the distribution $F(\alpha,\beta)$ will become Gaussian for large $N$, except
in the tails, where the linearized equation does not hold.

The Gaussian nature of the distribution also carries over to the finite time
average of $\alpha$,
\[
 \bar \alpha = \frac{1}{\tau} \int_0^{\tau} \alpha(t) dt .
\]
For large times $\tau$, the thermostat will spend a long time in the
neighborhood of the fixed point $(\alpha_0,\beta_0)$, so $\bar\alpha$ will also
be Gaussian distributed for large $N$, but, again, with deviations for large
fluctuations. This all is in accordance with what one would expect from a
central limit theorem.

Bonetto {\em et al.}\cite{Bonetto} find a Gaussian distribution for
$\bar\alpha$, but mention that the Gaussian was not what they expected, in
fact, that it couldn't be Gaussian, because it would give a kind of generalized
fluctuation-dissipation relation (see also \cite{EvansSearles}) when combined
with the fluctuation theorem of Gallavotti and Cohen\cite{GallavottiCohen}.

However, the deviations from the Gaussian nature at large fluctuations prohibit
this derivation of far-from-equilibrium fluctuation-dissipation relations. The
fluctuation theorem states that the probability $\pi(p)$ of finding that
$\bar\alpha$ is $p\alpha_{0}$, divided by the probability that it is
$-p\alpha_{0}$, satisfies for large $\tau$,
\[
  \frac{\pi(p)}{\pi(-p)} = \exp\left[ \tau d N
 \alpha_0 p \right].
\]
The result was found for Anosov systems, but has been found in numerical
simulations \cite{EvansCohenMorriss} and in Langevin equations\cite{Kurchan}
too, and seems to have a broader validity\cite{Crooks}. Combined with a
Gaussian form of $\pi$, one gets a relation between the variance and the mean
of the distribution, i.e.\ a kind of fluctuation-dissipation relation. The
variance can be linked to a correlation function and (generalized) Green-Kubo
formulas emerge\cite{Bonetto}. But for these to hold, the Gaussian character
has to be guaranteed for negative $\bar\alpha$ also, and the central limit
theorem, nor an extension of the analysis given here, could justify that.
Recently this was acknowledged by Searles and Evans\cite{SearlesEvans} and the
Green-Kubo relations far from equilibrium were refuted in their molecular
dynamics simulations.  This does not totally explain the results of Bonetto
{\em et al}\cite{Bonetto}, who get a Gaussian $\pi_{\tau}(p)$ for just 10
particles, the Gaussian clearly covering negative $p$ as well.  There are of
course uninvestigated prefactors in the widths of the Gaussian and in range of
validity.

Only when we are near equilibrium the fluctuation theorem and the central limit
theorem can be combined, provided we first take the limit of the external field
going to zero before we take the thermodynamic limit, even though we can take a
large number of particles throughout.  In such a scheme, the distribution
$\pi(p)$ is centered almost around zero, such that it always applies to some
negative $p$.

\section{Conclusions}

We have treated the Gaussian thermostat in a sheared system of short range
interacting particles at low density using kinetic equations. We found that in
the thermodynamic limit, the thermostat force becomes a friction force with a
constant friction coefficient. The value of this constant was shown to be
consistent with the requirement that the second moment of the one particle
distribution function is conserved. This conclusion did not depend on a
smallness of the shear rate, so it applies also far from equilibrium. The
constant friction force has been used in other work\cite{LeeDufty,Lutsko} and
the results from those should apply equally to Gaussian thermostatted systems,
in the thermodynamic limit.

We briefly discussed finite $N$ corrections. These give rise to fluctuations of
the friction coefficient around its mean, of the order of $1/\sqrt N$. The
fluctuations are close to Gaussian for large $N$, except for very large
fluctuations away from the mean.  Far-from-equilibrium Green-Kubo relations
rely in one derivation at least\cite{Bonetto} on the fluctuation
theorem\cite{GallavottiCohen}, which concerns large fluctuations. The breakdown
of such Green-Kubo relations recently found by Searles and
Evans\cite{SearlesEvans} has its origin in the deviations from the Gaussian
nature for large fluctuations. Near equilibrium one doesn't need the large
fluctuations.

\section*{Acknowledgements}

The author thanks Prof. H. van Beijeren, Prof. J.R.\ Dorfman and Prof. E.G.D.\
Cohen for valuable discussions. He acknowledges the hospitality and support of
the Institute of Science and Technology at the University of Maryland in
College Park and of the Rockefeller University in New York. The work reported
here was supported by FOM, SMC and by the NWO Priority Program Non-Linear
Systems, which are financially supported by the "Nederlandse Organisatie voor
Wetenschappelijk Onderzoek (NWO)".

\end{document}